# Electronic Origin for the Enhanced Thermoelectric Efficiency of Cu$_2$Se

Shucui Sun[a,1], Yiwei Li[b,1], Yujie Chen[a], Xiang Xu[a], Lu Kang[a], Jingsong Zhou[a], Wei Xia[c], Shuai Liu[c], Meixiao Wang[c,d], Juan Jiang[c], Aiji Liang[c], Ding Pei[b], Kunpeng Zhao[d], Pengfei Qiu[d], Xun Shi[d], Lidong Chen[d], Yanfeng Guo[c], Zhengguo Wang[e], Yan Zhang[e], Zhongkai Liu[c,f], Lexian Yang[a,g,*], Yulin Chen[a,b,c,f,*]

[a]*State Key Laboratory of Low Dimensional Quantum Physics, Department of Physics, Tsinghua University, Beijing 100084, China*
[b]*Department of Physics, Clarendon Laboratory, University of Oxford, Parks Road, Oxford OX1 3PU, UK*
[c]*School of Physical Science and Technology, ShanghaiTech University, Shanghai 201210, China*
[d]*State Key Laboratory of High Performance Ceramics and Superfine Microstructure, Shanghai Institute of Ceramics, Chinese Academy of Sciences, Shanghai 200050, China*
[e]*International Center for Quantum Materials, School of Physics, Peking University, Beijing 100871, China*
[f]*ShanghaiTech Laboratory for Topological Physics, Shanghai 200031, China.*
[g]*Frontier Science Center for Quantum Information, Beijing 100084, China*

[1]*These authors contributed equally to this work.*
*Email:\*lxyang@tsinghua.edu.cn,\*yulunchen@tsinghua.edu.cn*

**Thermoelectric materials (TMs) can uniquely convert waste heat into electricity, which provides a potential solution for the global energy crisis that is increasingly severe. Bulk Cu$_2$Se, with ionic conductivity of Cu ions, exhibits a significant enhancement of its thermoelectric figure of merit *zT* by a factor of ~3 near its structural transition around 400 K. Here, we show a systematic study of the electronic structure of Cu$_2$Se and its temperature evolution using high-resolution angle-resolved photoemission spectroscopy. Upon heating across the structural transition, the electronic states near the corner of the Brillouin zone gradually disappear, while the bands near the centre of Brillouin zone shift abruptly towards high binding energies and develop an energy gap. Interestingly, the observed band reconstruction well reproduces the temperature evolution of the Seebeck coefficient of Cu$_2$Se, providing an electronic origin for the drastic enhancement of the thermoelectric performance near 400 K. The current results not only bridge among structural phase transition, electronic structures and thermoelectric properties in a condensed matter system, but**

**also provide valuable insights into the search and design of new generation of thermoelectric materials.**

1. Introduction

Thermoelectric (TE) materials encapsulate unique capability to harvest waste heat and deliver electric energy, which provides broad and versatile application potential in energy, electronics and semiconductor industries [1-4]. Searching for TE materials with high TE efficiency, valued by the dimensionless figure of merit ($zT$), has become the central issue for the development of TE technology. By definition, $zT = \frac{\sigma S^2 T}{\kappa}$, which favors high electric conductivity $\sigma$, high Seebeck coefficient $S$ and low thermal conductivity $\kappa$ [5, 6]. Following this strategy, different TE materials have been discovered or created in various forms, such as nanostructured alloys of chalcogenides, filled skutterudites and clathrates, half-Heusler compounds, oxides, organic materials and ionic conductive materials [7-13]. Correspondingly, various mechanisms are proposed to explain the enhancement of the TE efficiency, including band convergence [14-16], Peierls distortion [17], impurity resonance [18, 19], modulation doping[20], lattice anharmonicity [21, 22], large electron effective mass[23], topological quantum effects and phonon-glass electron-crystal mechanism [24, 25].

Among various TE materials, superionic $Cu_2Se$, has attracted considerable research attention due to its simple chemical compositions and high $zT$ value (~ 1.5 at 1000 K, among the highest values in bulk materials) [12, 26-30]. It is widely believed that the superionic diffusion of $Cu^+$ at high temperatures strongly suppresses the lattice thermal conductivity, while the Se sub-lattice remains ordered and maintains the electric transport, which guarantees the high $zT$ value at high temperatures. Interestingly, a dramatic enhancement of $zT$ value by a factor of ~3 [26, 29, 31, 32] is observed across the structural phase transition from low-temperature α-phase to high-temperature β-phase near 400K [33], which is still mysterious and unsolved. In order to fully understand and improve its TE performance of $Cu_2Se$, it is essential to study both its electronic and structural properties. Nevertheless, previous experimental efforts mainly focus on the structural properties such as superionic behavior of Cu ions, while the other crucial ingredient in the system, the electrons, are largely ignored. Especially, the crucial electronic band structure is only investigated by theoretical calculations up to date [34, 35].

In this Letter, using high-resolution angle-resolved photoemission spectroscopy (ARPES), we systematically investigated the electronic structure of $Cu_2Se$ and its temperature evolution. Around the structural transition, we observed an abrupt formation of an energy gap at the $\bar{\Gamma}$ point near 400 K, together with a sudden change of the X-ray diffraction (XRD) peak position, suggesting a first-order nature of the α–β structural transition, which may be associated with a sudden change of lattice constants [33, 36] and drastic collective migrations of atoms [37]. On the other hand, the $\sqrt{3} \times \sqrt{3}$ reconstructed bands near $\bar{K}$ gradually disappear, which alludes a continuous order-disorder crossover of Cu ions [33] and is consistent with the continuous change in the transport measurements [29, 38]. Prominently, the observed band reconstruction well reproduces the drastic temperature evolution of the Seebeck coefficient across the phase transition. We thus conclude that the elevation of the $zT$ value in $Cu_2Se$ near 400 K is due to a synergetic effect of band-reconstruction enhanced Seebeck coefficient and the suppression of lattice thermal conductivity $\kappa$ by $Cu^+$ disordering. Our results present important insights into TE behaviour in $Cu_2Se$ by putting forward the electron degree of freedom, which will help search and design new generation of TE materials.

## 2. Materials and methods

*2.1 Sample synthesis.*

Single crystals of $Cu_2Se$ were synthesized by a melting-annealing-sintering process.[12, 39]. High-purity raw materials of Cu (Sigma Aldrich, 99.999%) and Se (Sigma Aldrich, 99.999%) were weighed out and loaded into a pyrolytic boron nitride crucible. Then the crucible was sealed in evacuated quartz tubes, which was slowly heated to 1423 K in 12 h, dwelled at this temperature for 24 h, slowly cooled to 1073 K in 50 h, and annealed at this temperature for 8 days. The obtained ingots were crushed into powders and then sintered at 753 K under a pressure of 65 MPa using a spark plasma sintering (Sumitomo SPS-2040) apparatus. Boron nitride was sprayed on the inner surfaces of graphite die to prevent currents going through the samples. Highly dense pellet samples with 10 mm in diameter and ~5 mm in thickness were finally obtained, and single crystals of $Cu_2Se$ with typical size of 0.5~1 mm for ARPES measurements were separated from the pellet samples.

*2.2 ARPES measurements.*

ARPES measurements were performed at beamline U09 of Shanghai Synchrotron Radiation Facility (SSRF), China, beamline BL13U of National Synchrotron Radiation Laboratory (NSRL), China, Surface/Interface Spectroscopy (SIS, X09LA) beamline at the Swiss Light Source (SLS) – Paul Scherrer Institute, Swiss and He-lamp based system in

Tsinghua University and Peking University, China. The samples were cleaved *in-situ* and measured under ultra-high vacuum below $6 \times 10^{-11}$ mbar. Data were collected with Scienta R4000 electron analyser at SSRF and NSRL, R2000 at SLS and DA30 analyser at Tsinghua and Peking University. The overall energy and angle resolutions were better than 20 meV and 0.2°, respectively.

*2.3 Scanning Tunneling Microscope (STM).*

STM measurements were performed at ShanghaiTech University, China. $Cu_2Se$ single crystals were cleaved in sample preparation chamber and transferred to STM chamber for measurement under ultra-high vacuum below $1.5 \times 10^{-11}$ Torr. Data were collected at room temperature using PtIr tips that were calibrated by silver islands grown on Si (111)-7×7 surface.

*2.4 Seebeck coefficient and electrical resistivity measurements.*

The Seebeck coefficient and electrical resistivity were measured using our newly developed system based on the thermal expansion equipment (Netzsch DIL 402C) [29]. The electrical resistivity was measured by the standard four-probe method. The Seebeck coefficient was calculated from the slope of the ΔV vs ΔT curve with a heating rate of 0.02 K/min, where ΔV and ΔT are the Seebeck voltage and temperature gradient, respectively. The Seebeck voltage ΔV was measured with a Keithley nanovoltmeter and further corrected for the thermopower of Pt. The temperature dependence of Seebeck coefficient was measured under a very small temperature step (around 0.1 K) between 320 K and 450 K.

**3. Results and discussion**

In the high temperature β-phase, Cu$_2$Se crystallizes in a face-centred cubic sublattice of Se atoms with superionic Cu ions diffusing inside [33, 40], which is isostructural with the topological insulator Ag$_2$Te [41]. Fig. 1a shows the anti-fluorite structure of β-Cu$_2$Se, which can be viewed as a quasi-layered structure along [111] direction (Fig. 1b). The Cu ions can randomly occupy different positions in the tetrahedron of Se atoms (Fig. 1a) [40, 42]. Below the transition temperature, Cu$_2$Se transforms into the monoclinic α-phase [33, 40]. Although the accurate crystal structure of α-Cu$_2$Se is still under debate, it is commonly accepted that there exist alternating Cu-deficient and Cu-rich Se-Cu-Se layers in the crystal [33, 40], as schematically shown in Fig. 1c. Such a layered structure is evidenced by the height of the terrace measured by scanning tunnelling microscopy (Fig. 1d). The topography of the topmost atoms shows dumbbell-like dimerization (Fig. 1e), which is in excellent agreement with the surface Cu$^+$ arrangement in Fig. 1c. This dimerization supports a $\sqrt{3} \times \sqrt{3}$ reconstruction of Cu ions (Fig. S2), which is also suggested by previous XRD measurements [33, 40]. The Resistivity as a function of temperature shows a clear anomaly near 400 K, verifying the structural transition in our samples (Fig. 1f). Further sample characterization can be found in the Support Materials (Fig. S1).

The electronic structure of α-Cu$_2$Se is investigated by ARPES at 15 K. Fig. 2a shows the three-dimensional BZ of α-Cu$_2$Se and (001) surface projection. Due to the rearrangement of Cu ions, the periodicity in the momentum space is reduced to half along $k_z$ and 1/3 in $k_x$-$k_y$ plane compared to β phase, consistent with the structural change as shown in Fig. 1. The Fermi surface map in $k_x$-$k_y$ plane in Fig. 2b shows a clear hexagonal structure near both the $\bar{\Gamma}$ and $\bar{K}$ points, which are contributed by the hole bands shown along $\bar{K} - \bar{\Gamma} - \bar{K}$ (Fig. 2c). The bands near $\bar{K}$ are back-folded from $\bar{\Gamma}$ due to √3 × √3 Cu⁺ ordering (Fig. 1e). Photon energy dependent measurement shows clear $k_z$ periodicity at $E_F$ (Fig. 2d) suggesting the bulk origin of the measured bands. The fine FS map near Γ shows a circular hole pocket and a warped hexagonal hole pocket that evolves to a hexagram at high binding energies. 5 bands are resolved in an energy range of about 1.2 eV below $E_F$, as shown in Fig. 2f. The weak intensity near the top of the δ band may be due to the $k_z$ contamination. Near the Z point, the FS is dominated by a triangular feature that evolves to a complex structure consisting of three overlaid ellipses (Fig. 2g). The η band is out of the detected energy range near Z due to its strong $k_z$ dispersion (Fig. S4). Our measurement shows qualitative agreement with previous theoretical calculation of the electronic structure of α-Cu$_2$Se with a similar crystal structure shown in Fig. 1c [26].

The intriguing α–β transition of Cu$_2$Se is investigated by the temperature evolution of both crystal and electronic structure. Our temperature dependent single crystal X-ray diffraction (XRD) measurements along [001] (or [111] in β-phase) show a clear shift of XRD scattering peak near 26.3° together with the disappearance of the scattering peaks near 12.9° and 39.7° above 400 K (Fig. 3a and Fig. S3), suggesting a sudden change of the crystal structure along the interlayer direction [43]. Interestingly, we observe a strong negative thermal expansion along *c* prior to the structural transition already starts from 300 K, which is attributed to the gradual reordering of Cu ions with temperature [33].

While the sudden change of the XRD pattern alludes a first-order character of the structural transition near 400 K, the band structure exhibits a more complex temperature evolution. Fig. 3b and Fig. 3c compare the FS measured over multiple BZs at 10 K and 396

K, respectively. The electronic states near $\bar{K}$ disappear at 396 K, consistent with the calculated electronic structure of β-Cu$_2$Se [44]. With increasing temperature, the reordering of Cu ions as indicated by the negative thermal expansion reduces the √3 × √3 superstructure and suppresses the spectral intensity near $\bar{K}$. Interestingly, the spectral intensity near $\bar{K}$ shows a continuous reduction to zero starting from far below 400 K, as shown in Fig. 3d and Fig. S6, consistent with the ordering-disordering crossover of Cu ions [29, 33, 38]. On the other hand, near the $\bar{\Gamma}$ point, the hole bands shift towards high binding energies in a narrow temperature range from 390 K to 400 K, and finally develop a band gap of about 150 meV near $\bar{\Gamma}$ (Fig. 3e-h), suggesting the p-type semiconducting nature of β-Cu$_2$Se. In contrast to the continuous decrease of the intensity near $\bar{K}$ (Fig.3d), both the intensity drop and the band shift near $\bar{\Gamma}$ become drastic from 390 K (Fig. 3g, h), which may be associated with a sudden acceleration of Cu$^+$ hopping [37] and the subsequent structural transition.

Temperature dependent measurements on both band and crystal structures indicate that the α–β transition consists of a first-order structural transition and a much slower continuous process. Upon cooling below 400 K, the crystal undergoes a sudden expansion along $c$ and the Cu ions are almost "frozen" with suppressed hopping rates through the first-order transition [36, 37], leading to the band gap close at $\bar{\Gamma}$. In a much larger temperature range, the Cu ions continuously migrate along $c$ and order in $ab$ plane [33], forming the back-folded bands near the $\bar{K}$ point. The proposed scenario may provide a microscopic interpretation for the exotic temperature evolution of the thermal conductivity of Cu$_2$Se [29, 32, 38], which exhibits a drastic and continuous response to temperature change in different temperature ranges.

The observed band reconstruction indicates a change of the chemical potential in the system during the structural transition, which can play a crucial role in the TE properties of $Cu_2Se$ that are sensitive to the density of states near $E_F$ [27]. Indeed, when the system is in the disordered regime (short carrier lifetime and mean-free-path), the Seebeck coefficient $S$ of a p-type semiconductor can be evaluated experimentally as [45]:

$$S = \frac{2\pi^2 k_B^2 T}{3e} \frac{d \ln N(E)}{dE}\bigg|_{E=E_F} \quad (1)$$

where $k_B$, $e$ and $N(E)$ are Boltzmann constant, electron charge and carrier density of states, respectively. It is worth noting that a rough approximation $\sigma \propto \{N(E_F)\}^2$ was adopted in equation 1, where $\sigma$ denotes the electrical conductivity [45].

In Fig. 4, we approximate the density of states near $E_F$ with ARPES intensity $I(E)$ to simulate the Seebeck coefficient. Since $S$ is determined mainly by the electronic properties in a narrow energy window (a few $k_B T$) around $E_F$, we average the value of $\frac{2\pi^2 k_B^2 T}{3e} \frac{d \ln I(E)}{dE}$ over an energy window of 120 meV (~$4k_B T$) near $E_F$ and show the result as a function of temperature in Fig. 4 (the grey curve). The approximation based on ARPES data qualitatively reproduces the evolution of the Seebeck coefficient across the α−β transition independently measured by the transport experiment (the blue curve in Fig. 4), confirming the dominant role of the electronic structure in the Seebeck coefficient of $Cu_2Se$.

In order to further understand the influence of the electronic structure on the Seebeck coefficient, we simulate the temperature evolution the Seebeck coefficient according to equation (1) based on a simple band-shift model. In this model, the band structure is approximated by a parabolic dispersion with its valence top energy $E_F$ - $E_V$ extracted from our ARPES measurements (Fig. 3h and Fig. S7). A Lorentzian broadening of 200 meV (FWHM) is applied on the band structure to simulate the disordered effect of finite carrier lifetime and mean-free-path. Prominently, as shown by the orange curve in Fig. 4, our simulation based on equation (1) excellently reproduces the temperature evolution of the Seebeck coefficient near 400 K, indicating a dominant contribution of the band reconstruction in the enhanced Seebeck coefficient of $Cu_2Se$.

Our finding of band-reconstruction enhanced Seebeck coefficient is supported by the strong influence of the chemical potential on the Seebeck coefficient of $Cu_2Se$ calculated using the Boltzmann transport equation [27]. However, as an oversimplified model, quantitative differences between the simulated and transport measured Seebeck coefficient are still observed in Fig. 4. On the one hand, equation (1) was based on a rough approximation [45]. On the other hand, other effects in the band structure, such as the multi-band nature, the change of electron effective mass, temperature dependent scattering rate and the non-rigid band shift across the structural transition as revealed by the band structure calculations can likewise contribute to the Seebeck coefficient, which requires further experimental and theoretical investigation.

## 4. Conclusion

We present a comprehensive investigation of $Cu_2Se$ using high-resolution ARPES and XRD. Our results reveal the disentangled structural evolution of Cu and Se sub-lattices during the structural transition. We conclude that the band reconstruction plays a dominant role in the enhancement of the Seebeck coefficient, which together with the suppression of the lattice thermal conductivity due to $Cu^+$ disordering, significantly improve the TE performance of Cu$_2$Se near the α−β structural transition. Our results provide important insights into not only the peculiar electronic and structural phase transitions in Cu$_2$Se, but also the excellent TE efficiency of Cu$_2$Se, which will further shed light on its application in the TE devices.

**Acknowledgments**

We thank Zhe Sun, Yuliang Li, Yaobo Huang, Hong Ding, Junzhang Ma, and Ming Shi for experimental support. This work was supported by the National Natural Science Foundation of China (Grant No. 11774190, No. 11674229, No. 11634009 and No. 11874264), the National Key R&D program of China (Grant No. 2017YFA0304600, 2017YFA0305400 and 2017YFA0402900), EPSRC Platform Grant (Grant No. EP/M020517/1). Yanfeng Guo acknowledges the support from the Natural Science Foundation of Shanghai (No. 17ZR1443300). Lexian Yang and Yulin Chen acknowledge the support from Tsinghua University Initiative Scientific Research Program.

**Conflict of interest**

The authors declare that they have no conflict of interest.

**Author contributions**

Shucui Sun and Yiwei Li contributed equally to this work. Yulin Chen and Lexian Yang conceived the experiments. Shucui Sun and Yiwei Li carried out ARPES experiments with the assistance of Yujie Chen, Xiang Xu, Lu Kang, Jingsong Zhou, Juan Jiang, Aiji Liang, Ding Pei, Zhengguo Wang, Yan Zhang, and Zhongkai Liu. Shuai Liu and Meixiao Wang performed STM measurements. Wei Xia, Kunpeng Zhao, Pengfei Qiu, Xun Shi, Lidong Chen and Yanfeng Guo synthesized Cu$_2$Se single crystals and measured the Seebeck coefficient. All authors contributed to the scientific planning and discussions.

**Supplementary materials**

Supplementary materials to this article can be found online or from the authors.

**Figures and legends**

**Fig. 1.** Basic characterization of Cu$_2$Se crystals. (a) Schematic illustration of the crystal structure of β-Cu$_2$Se. (b) The side view along [111] of the crystal structure of β-Cu$_2$Se. (c) Schematic illustration of the crystal structure of α-Cu$_2$Se. (d) Large-scale surface topography of Cu$_2$Se measured with sample bias of 0.8V and tunneling current of 250 pA. The inset shows the line profile along the blue line. (e) Surface topography with atomic resolution showing the √3 × √3 superstructure with sample bias of 10mV and tunneling current of 250 pA. The STM data was collected at room temperature. (f) Resistivity of Cu$_2$Se as a function of temperature.

**Fig. 2.** Fermi surface and band dispersion of α-Cu$_2$Se. (a) Three-dimensional Brillouin zone (BZ) of α-Cu$_2$Se and its surface projection. The red and blue lines indicate the surface BZ without and with taking √3 × √3 Cu$^+$ ordering into account, respectively. (b) Fermi surface (FS) in $k_x$-$k_y$ plane obtained by integrating ARPES intensity in an energy window of 20 meV around $E_F$. (c) The band dispersion along $\bar{K} - \bar{\Gamma} - \bar{K}$ showing hole bands near boh $\bar{\Gamma}$ and $\bar{K}$. (d) FS in $k_y$-$k_z$ plane. (e) Constant energy contours in $k_x$-$k_y$ plane near the Γ point. (f) Band dispersion near the Γ point. (g,h) the same as (e,f) but near the Z point. Data in panels (e) and (f) were collected at 50 eV, while data in panels (g) and (h) were collected at 62 eV.

**Fig. 3.** Structural transition and band reconstruction across the phase transition. (a) Temperature dependent XRD measurement. (b,c) FS measured at 20 K and 396 K, respectively. (d) Temperature evolution of ARPES intensity near the $\bar{K}$ point. (e) The comparison between the dispersion near $\bar{\Gamma}$ measured at 306 K and 450 K. (f) False-color plot of temperature evolution of energy distribution curve near $\bar{\Gamma}$. (g) Temperature evolution of ARPES intensity near $\bar{\Gamma}$

integrated in an energy window of 100 meV at -0.1 eV. (h) Temperature dependent band shift near $\bar{\Gamma}$.

**Fig. 4.** Comparison between the measured Seebeck coefficient (blue curve), calculated Seebeck coefficient with ARPES intensity (grey curve), and simulated Seebeck Coefficient using the simple model described in the main text (orange curve) as a function of temperature.

# Figure 1

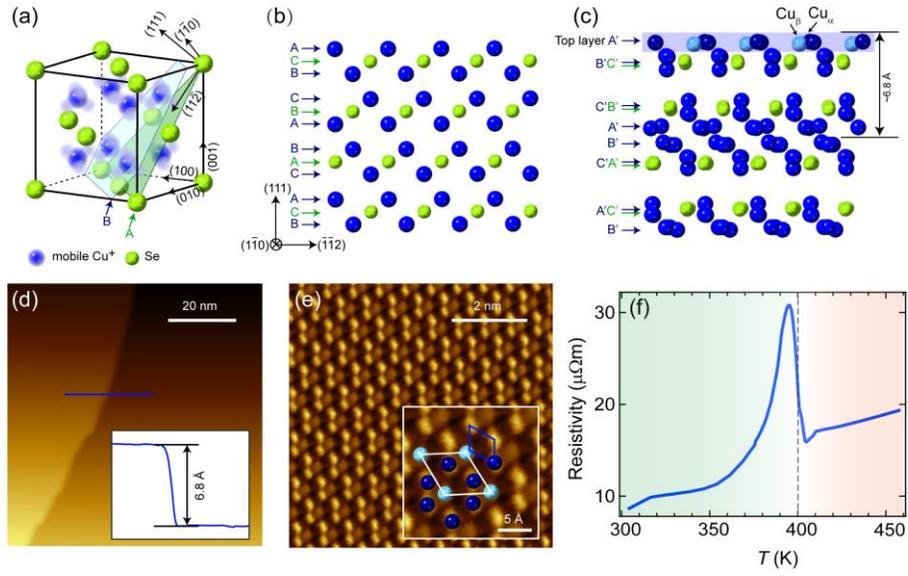

# Figure 2

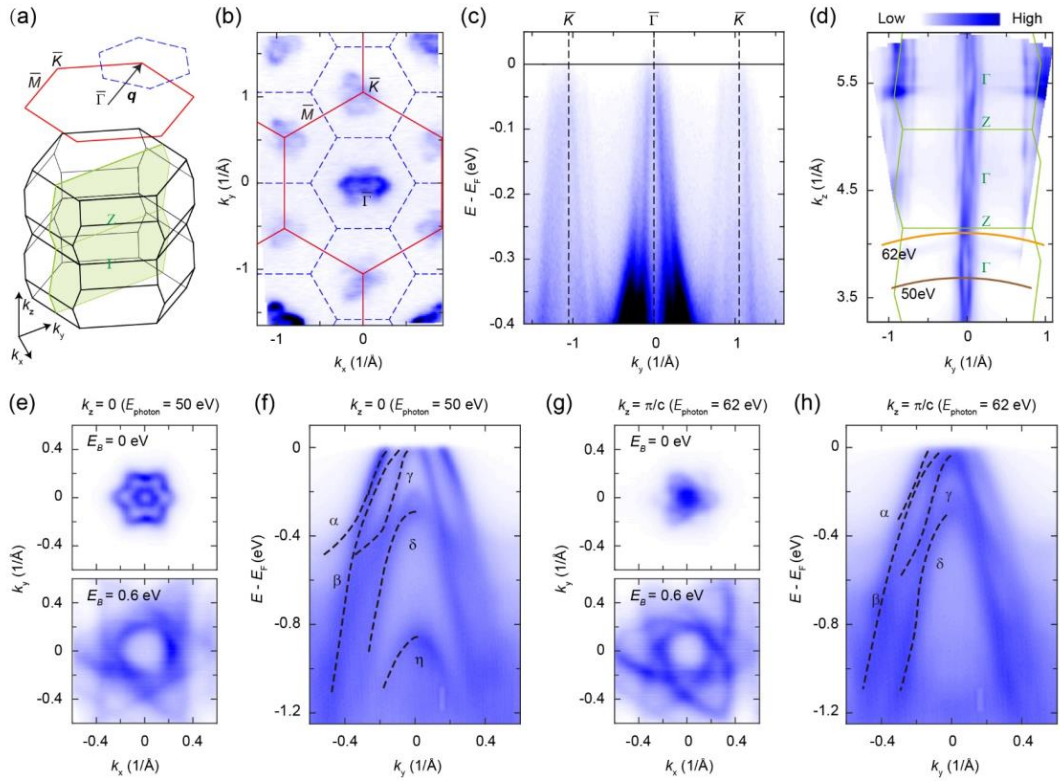

**Figure 3**

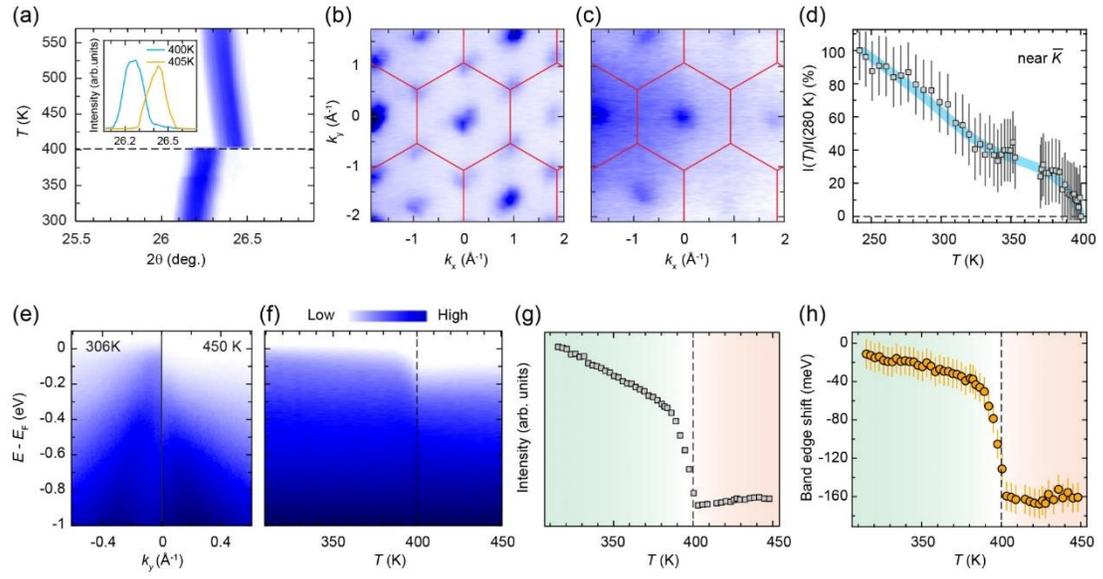

**Figure 4**

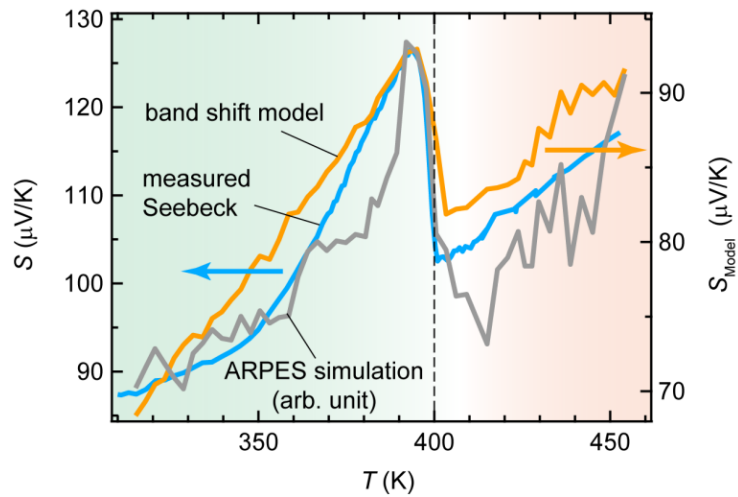